\documentclass[final,5p,times,twocolumn]{elsarticle}

\usepackage{graphicx,epstopdf}
\usepackage[usenames,dvipsnames]{color}
\usepackage{color}
\usepackage{amssymb}
\usepackage{lineno}

\newcommand{\ele}{\ensuremath{e^{+}}}
\newcommand{\pos}{\ensuremath{e^{-}}}
\newcommand{\eleca}{\texttt{EleCa}}
\newcommand{\hermes}{\texttt{HERMES}}

\journal{Astroparticle Physics}

\begin{document}

\begin{frontmatter}

\title{Propagation of extragalactic photons at ultra-high energy with the EleCa code}

\author[cor1,cor2,fn1]{Mariangela Settimo}
\address[cor1]{Laboratoire de Physique Nucl\'eaire et de Hautes Energies (LPNHE), Universit\'es Paris 6 et Paris 7, CNRS-IN2P3, Paris, France. \\}
\address[cor2]{University of Siegen, Germany}
\fntext[fn1]{\emph{email: mariangela.settimo@lpnhe.in2p3.fr}} 
\author[cor3,cor4]{Manlio De Domenico}
\address[cor3]{Departament d'Enginyeria Inform\`atica i Matem\`atiques, Universitat Rovira i Virgili, Tarragona, Spain}
\address[cor4]{Laboratory of Complex Systems, Scuola Superiore di Catania, Italy}

\begin{abstract}
Ultra-high energy (UHE) photons play an important role as an independent probe of the photo-pion production mechanism by UHE cosmic rays. Their observation, or non-observation, may constrain astrophysical scenarios for the origin of UHECRs and help to understand the nature of the flux suppression observed by several experiments at energies above $10^{19.5}$~eV. The interaction length of UHE photons above $10^{17}$~eV ranges from a few hundred kpc up to tens of Mpc. Interactions with the extragalactic background radiation initiate the development of electromagnetic cascades which affect the fluxes of photons observed at Earth. The interpretation of the current experimental results rely on the simulations of the UHE photon propagation.   
In this paper, we present the novel Monte Carlo code ``\eleca'' to simulate the \emph{Ele}ctromagnetic \emph{Ca}scading initiated by high-energy photons and electrons. 

We provide an estimation of the survival probability for photons inducing electromagnetic cascades as a function of their distance from the observer and we calculate the distances within which we expect to observe UHE photons with energy between $10^{17}$ and $10^{19}$~eV. Furthermore, the flux of GZK photons at Earth is investigated in several astrophysical scenarios where we vary both the injection spectrum and composition, and the intensity of the intervening extragalactic magnetic field.
Although the photon propagation depends on several astrophysical factors, our numerical predictions combined with future experimental observations (or non-observations) of UHE photons in the energy range between $10^{17.5}$~eV and $10^{20}$~eV can help to constrain these scenarios.
\end{abstract}


\end{frontmatter}
%\linenumbers

\section{Introduction}

Ultra-high energy (UHE) photons are expected to be produced in the interactions of UHE cosmic rays with matter, e.g. with gas around a source, or with the extragalactic background radiation (EBR). In particular, photons can be created from the decay of neutral pions produced in the interaction of UHE nuclei with the EBR photons, for instance, because of the Greisen-Zatsepin-Kuz'min (GZK) effect~\cite{Greisen:1966jv,Zatsepin:1966jv}. 
Recent measurements of the UHECR energy spectrum~\cite{Zhang:2008zze,Settimo:2012zz} have confirmed a flux suppression above 50 EeV (1\,EeV\,=\,$10^{18}$~eV) compatible with the expectation from GZK effect, although other alternatives (e.g., a physical limit to the maximum acceleration at the source) can not be ruled out. Hence, the observation of UHE photons would be an independent evidence for the existence of the GZK effect, providing hints on the nature of UHECRs, on astrophysical sources and on environmental conditions. 
A large fraction of UHE photons are also expected, from the decay or the annihilation of supermassive particles~\cite{Gelmini:2007sf}, in some exotic (top-down) models for UHECR acceleration. No UHE photons have been observed so far and upper limits have been placed on their flux~\cite{Settimo:2011zz,AgasaLimits,YakutskLimits,TelescopeArray}, partly disfavoring exotic models.
Current experimental results on the chemical composition of UHE cosmic rays~\cite{Abraham:2010yv,Abbasi:2009nf,Tameda:2011zz,Dedenko:2012ws,massWG2012} suggest a mixed composition from light to heavier nuclei at the highest energies. As will be discussed in the next sections, the expected flux of photons depends, among other parameters, on the chemical composition of CRs at their CRs sources. The observation, or non-observation, of photons would thus help to clarify the nature of the flux suppression observed at the highest energies and eventually  provide additional hints to the understanding of the chemical composition of UHECRs. 
 The estimate of the expected fluxes of photons is thus an important aspect to improve. This is particularly true if one takes into account the large variation of photon fluxes predicted at Earth by different simulators. Several predictions of photon fractions have been made in the past (see for example~\cite{Gelmini:2007sf,Ahlers:2011sd,Hooper:2006tn}).  Most of the adopted simulators are based on the solution of transport equations, or on semi-analytical approaches, and only a few of them are publicly available. For instance, among the publicly available codes, the Monte Carlo approach is adopted in~\cite{Kachelriess:2011bi} where, however, double and triple pair productions are not included. We will show in the next section that the latter is dominant in the energy region above 10$^{17}$~eV. 
 On the other hand, the solution of transport equations is the fundamental ingredient of the CRpropa code~\cite{Armengaud:2006fx}, allowing one to follow the propagation of photons at higher energies.

For these reasons, we present here a novel Monte Carlo simulation code to reproduce more realistically the propagation of extragalactic photons and their cascade development. This \emph{stand-alone} code, named \eleca$\,$(\emph{El}ectromagnetic \emph{Ca}scading), is based on a pure Monte Carlo approach and has been designed to propagate primary photons produced either by candidate sources or produced as secondary particles during the propagation of UHE nuclei and their interactions with relic photons of extragalactic background radiation. Our code is written in C++ with a highly modular structure which simplifies the interface with existing codes developed to simulate the propagation of UHECRs. 

In this paper, we describe the structure of \eleca, showing its basic features and some applications relevant to the understanding of the propagation of photons and nuclei. 
We also estimate the source distances within which we expect to observe UHE photons with energy between $10^{17}$ and $10^{19}$~eV and we show some applications of astrophysical interest.

The paper is organized as follows. The details of the interaction processes and their implementation in \eleca$\,$ are given in Section~\ref{sect:physics}. 
In the next section we discuss some physical results and applications, including (i) an estimate of the survival probability for photons inducing electromagnetic cascades as a function of their distance from the observer (section~\ref{sect:maxdist}), (ii) the prediction of the fluxes at Earth for GZK photons (section~\ref{sect:fluxes}) and (iii) the calculation of the corresponding horizons (section~\ref{sect:horizons}) in some representative astrophysical scenarios using the most recent energy spectrum measurement~\cite{Auger:icrc2013} for normalization.

\section{Propagation of photons}\label{sect:physics}

Ultra-high energy photons, with energy ranging from $10^{17}$~eV to $10^{18}$~eV, undergo interactions with the diffuse background radiation while travelling from their sources to the Earth.
In the energy range of interest in this work, the dominant energy-loss processes for UHE photons are Pair Production (PP) and Double Pair Production (DPP), responsible for the production of secondary electrons and positrons (hereafter generally referred as electrons). Electrons are cooled via Inverse Compton Scattering (ICS), $e\gamma_b\rightarrow e\gamma$, or through Triplet Pair Production (TPP), $e\gamma_b\rightarrow e\ele\pos$.  Moreover, the development of electromagnetic cascades depends on the strength of the intervening extragalactic magnetic fields (EGMFs). For sufficiently intense EGMFs, electrons in the cascade will lose most of their energy by synchrotron radiation and the cascade development will stop when the synchrotron cooling time scale becomes smaller than the ICS interaction mean free path. Here, it is worth noting that the influence of magnetic fields with strength of the order of $\mu$G, occurring around the cosmic
large-scale structure, is not accounted for because the production of secondary $e^{+}/e^{-}$ is supposed to occur far from sources. 

\begin{figure}[!t]
\hspace{-0.4cm}
\includegraphics[width=0.50\textwidth]{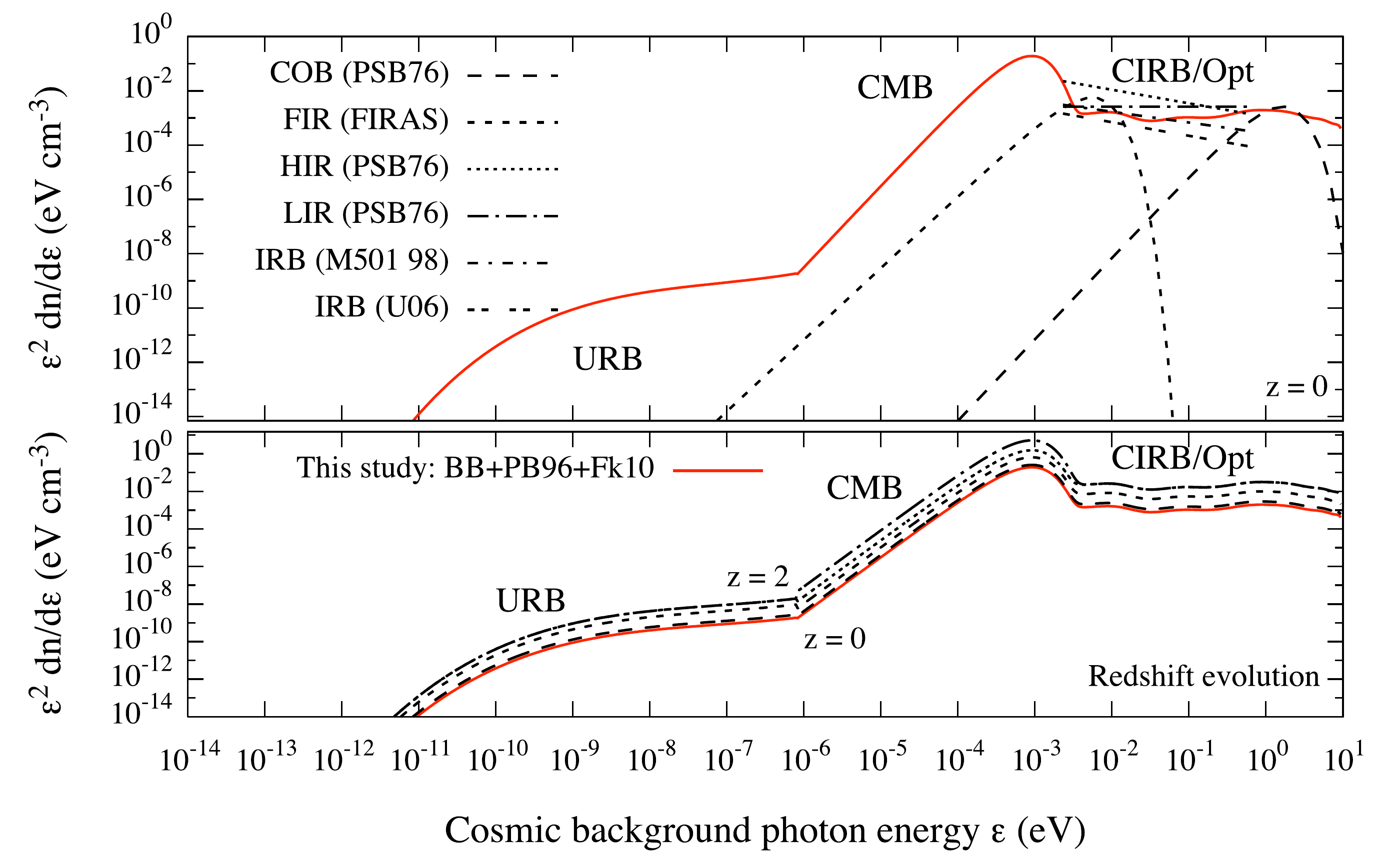}
\caption{Top panel:  different parameterizations of the number density, $n(\epsilon)$, of the extragalactic background radiation (EBR) as a function of relic photon energy. The red solid line indicates the EBR parameterization included in this study.  The other parameterizations, shown for reference, are taken from PSB76 \cite{puget1976photonuclear}, FIRAS \cite{fixsen2011probing}, ER98 \cite{epele1998propagation}, Mkn501-98 \cite{funk1998upper}, U06 \cite{uryson2006ultra}. Photon energy is considered in the laboratory frame. Bottom panel: evolution of EBR for different values of redshift $z$ ranging from 0 to 2.}
\label{fig:EBR}
\end{figure}

Some models of EBR are shown in the top panel of Fig.\,\ref{fig:EBR}, as a function of the relic photon energy $\epsilon$ in the laboratory frame. The red solid line indicates the EBR spectrum adopted in this work. In particular, the black-body model with temperature $T_{0}\simeq 2.725$~K is adopted for the cosmic microwave background (CMB) while the semi-analytical ``model D'' in~\cite{finke2010modeling} is adopted for the infrared/optical background (CIOB). The model proposed in~\cite{Protheroe:1996si} is adopted for the universal radio background (URB).
For the sake of completeness, we also show the models for COB (PSB76), for lower and higher infrared radiation (LIR and HIR, respectively) \cite{puget1976photonuclear}, and other infrared background models, derived from theoretical arguments or experimental observations \cite{epele1998propagation,funk1998upper,uryson2006ultra,fixsen2011probing}. The bottom panel of the same figure shows the evolution with redshift for different values of $z$, ranging from 0 to 2. 

The propagation of UHECRs in the Universe is characterized by the interaction length $\lambda$, depending, in general, on the interaction process, the background radiation, the redshift evolution and the particle type. The interaction length provides us with an estimation of the average distance travelled by a UHECR before undergoing interactions. In the case of UHE photons, for a given process with cross section $\sigma$, the interaction length $\lambda$ at present time can be calculated by 

\begin{equation}
\lambda^{-1} = \frac{1}{8\beta E^2}\int^\infty_{\epsilon_{min}}\frac{n(\epsilon)}{\epsilon^2}\int^{s_{max}}_{s_{min}} \sigma(s)(\,s\, -\, m^2c^4)\,ds\,d\epsilon
\label{eq:generalLambda}
\end{equation}
with $n(\epsilon)$ the density of background photons with energy $\epsilon$ in the observer's rest frame, $s$ the invariant energy of the center of mass, $m$ the mass of the incident particle and $\beta$c the velocity of the incident particle. The integration limits also depend on the process and the background radiation. For each process, we derived  $\lambda$ by solving numerically Eq.~(\ref{eq:generalLambda}) and we estimated it as a function of energy (up to $10^{23}$~eV) and for different background radiation. The obtained values are tabulated and successively used to speed up the cascade propagation, whereas the redshift evolution of $\lambda$ is estimated case-by-case during the simulation, depending on the background radiation involved in the interaction.

The total PP cross section for a photon with energy $E$ scattering off a soft-photon with energy $\epsilon$ is given by~\cite{Lee:1996es}, being proportional to the Thompson cross section $\sigma_T$ and to the $\beta$ of the incident particle.

\begin{figure}[!t]
\begin{center}
\includegraphics[width=0.42\textwidth]{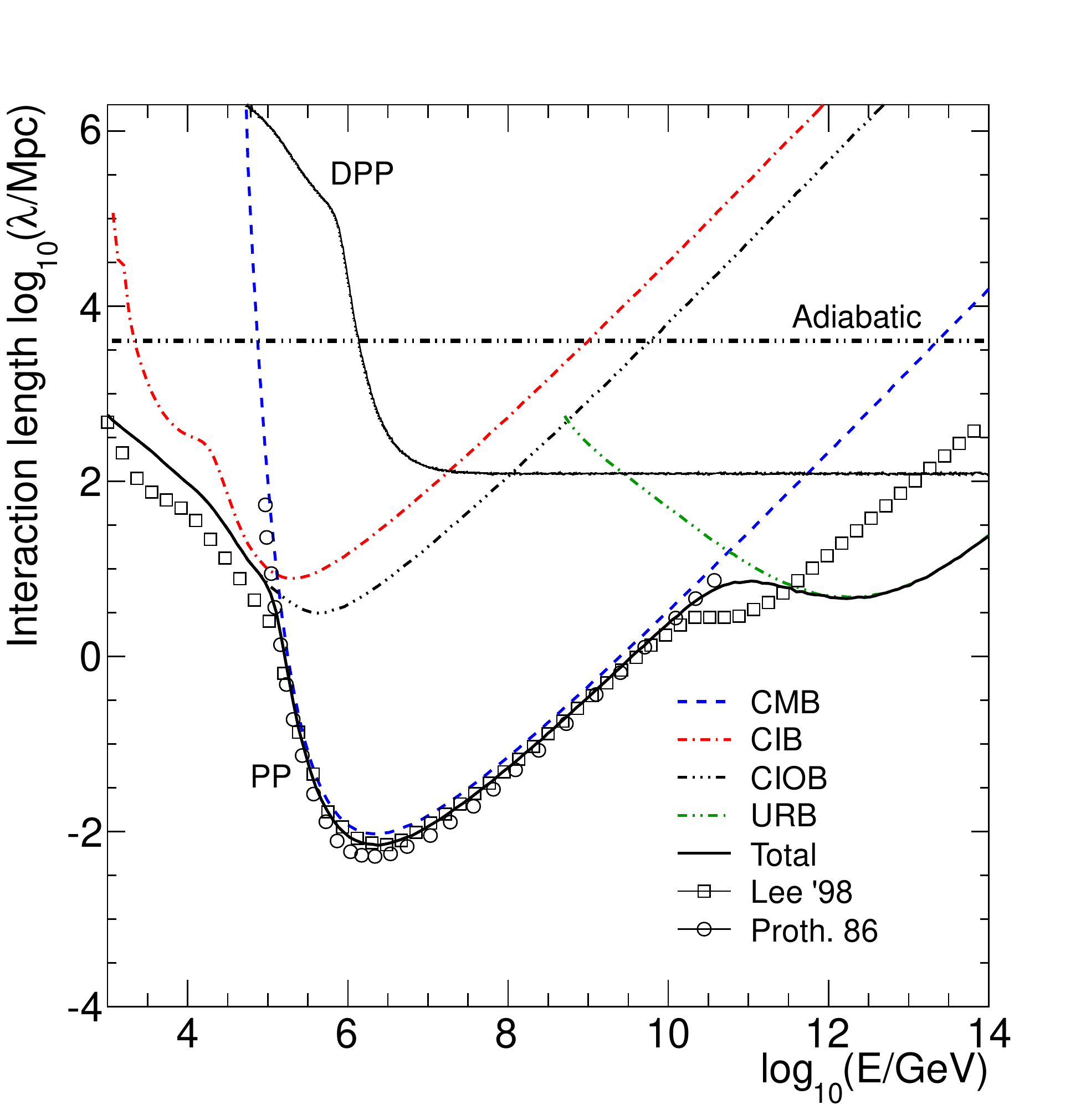}
\includegraphics[width=0.42\textwidth]{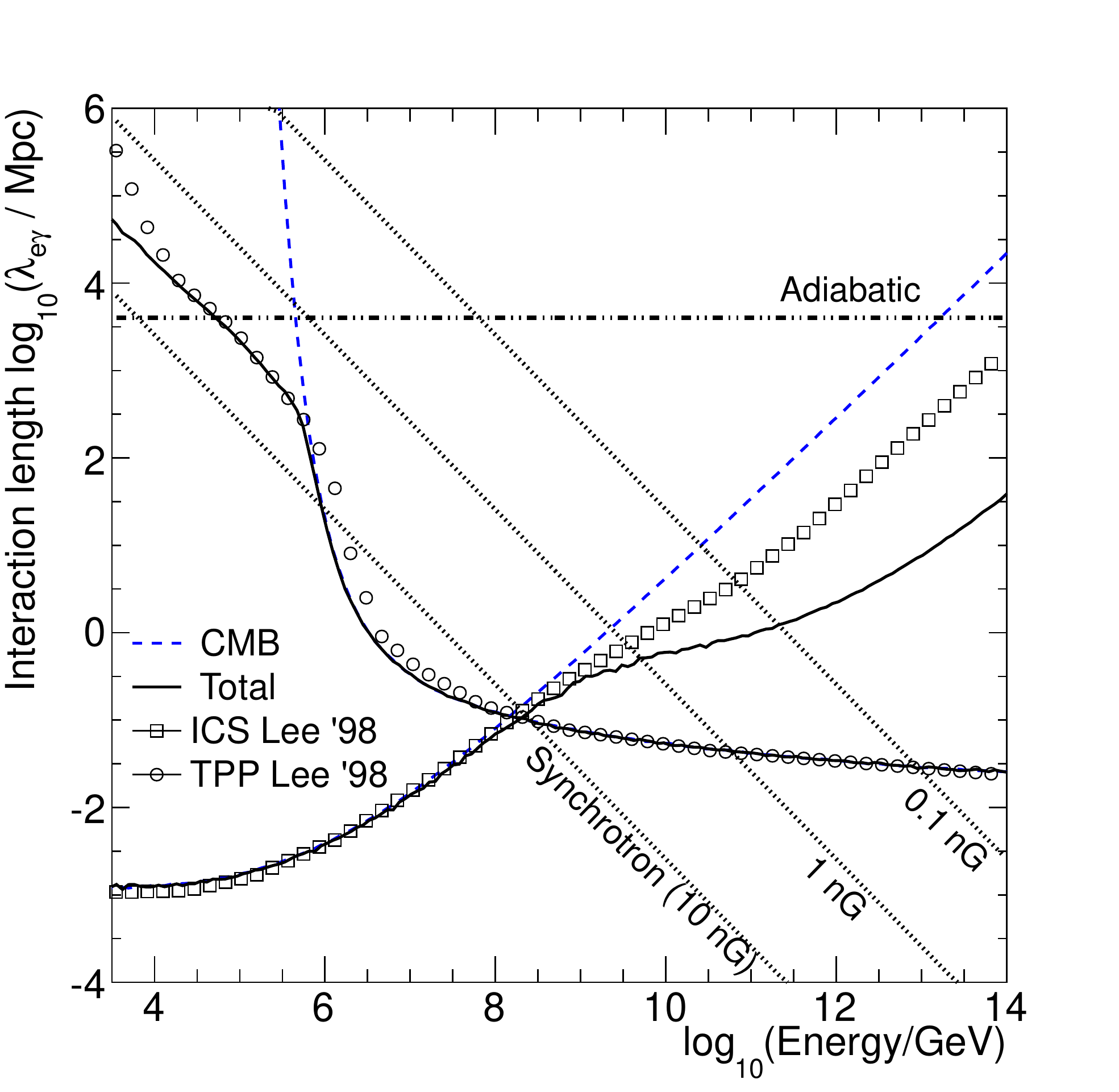}
\end{center}
\caption{Mean interaction lengths for photons (top) and electrons/positrons (bottom) for the whole EBR spectrum relevant for this study (solid). Dashed lines are given for each individual contributions to the EBR (see text).
The interaction length corresponding to the adiabatic and synchrotron energy losses are also shown as a dot-dashed and dotted lines respectively. }
\label{fig:lambda}
\end{figure}

The interaction length for the Double Pair Production (DPP), $\gamma\gamma_b\longrightarrow \ele\pos\ele\pos$, is derived by considering the corresponding cross section in the approximation reported in~\cite{Demidov:2008az,Brown:1973}. The process is subdominant over the full energy range if the PP is calculated on a realistic radio background. However, it is worth noting that the attenuation of the electromagnetic shower is dominated by TPP on URB above $10^{19}$~eV.

The interaction length at the present epoch ($z = 0$) for PP and DPP processes is shown in Fig.~\ref{fig:lambda} (top). The contribution of the whole EBR (solid line), and of each EBR component separately (dashed), are shown to emphasize the influence of each background radiation on the mean free path. As evident from the figure, photons with energy below tens of TeV, can propagate without interacting for distances of hundreds of Mpc, cooling down just because of adiabatic energy losses. On the other hand, the Universe becomes opaque to photons with energy of a few hundreds or thousands of TeV, getting more transparent at energies in the EeV range, where the radio background is the main responsible for energy-loss processes. Results assuming different EBL models, as in~\cite{Lee:1996es,Protheroe:1986},  are also shown for comparison: the different models assumed for the IRB and URB, are responsible for the observed differences at very low and high energies, respectively. 

In constrast to other types of interactions, the Inverse Compton Scattering (ICS) is a well known process occurring without threshold. 
It is worth noting that at low energy (namely below $\sim$1~TeV) $\lambda_{ICS}$ is constant because, for $s<<4\,m_{e}$, the cross section $\sigma_{ICS}\,=\,\sigma_T$. 
Above a few $10^{17}$~eV, the dominant process~\cite{Mastichiadis:1985,Mastichiadis:1991} is the TPP. Given the production of 3 electrons, even in the lower energy region, the contribution of TPP may be especially important in the presence of strong magnetic fields, because of the synchrotron energy loss affecting charged particles.
The interaction lengths of the two processes are shown in Fig.~\ref{fig:lambda} (bottom) for the whole EBR spectrum relevant for this study and for each component separately. The interaction lengths derived in previous works are also provided for comparison, leading to the same considerations already mentioned in the case of the PP and DPP. 

For the ICS, the energy of the secondary particles is derived according to the differential cross section~\cite{Coppi:1990}. For the TPP we used the approximation given in~\cite{Mastichiadis:1991,Lee:1996es} for the leading electron, while the leftover energy is equally shared between the remaining pair. 

Other sub-dominant processes (e.g. muon pair production on background photons, pair production in magnetic fields or on atoms, ions and free electrons) are thus neglected in this work. 

Finally, electron and photons adiabatically lose their energy because of the expansion of the Universe, while electrons might additionally lose energy because of the emission of synchrotron radiation while traversing EGMFs. 
 The energy-loss lengths corresponding to the average synchrotron radiation emission for \ele/\pos~\cite{Stanev:802298}, is also shown in Fig.~\ref{fig:lambda} (bottom), for three different intensities of the magnetic field. 
Deflections due to the extragalactic magnetic field are taken into account by assuming the ``small angle'' approximation~\cite{Hooper:2006tn}, which is valid for magnetic fields with strength smaller than a few nG at the energy scale of interest in this paper.
 The adiabatic energy loss rate for the propagation in a Friedmann Universe is given by:
\begin{eqnarray}
\beta_{rsh}(z)= H_{0}[\Omega_{M}(1+z)^{3} + \Omega_{\Lambda} + (1-\Omega_{M}-\Omega_{\Lambda})(1+z)^{2} ]^{\frac{1}{2}},\nonumber
\end{eqnarray}
where $H_{0}$ indicates the Hubble parameter at the present epoch, $\Omega_{M}$ and $\Omega_{\Lambda}$ are density parameters accounting for the matter and the energy in the Universe, respectively. 
In the case of synchrotron emission, we adopt a simplified average behaviour given in~\cite{Stanev:802298}. Within this approximation, only the strength of the magnetic field is taken into account. 

Summarizing, \eleca\, simulates the propagation $\gamma$, \ele and \pos, i.e., their interactions with relic photons and the subsequent electromagnetic cascading processes, with a given energy and initial position according to some standard configuration (e.g., power-law spectra, monochromatic sources) or from a list of particles provided externally by the user. 
Apart for the adiabatic and the synchrotron energy losses, all the interactions, including DPP and TPP up to the highest energies, are treated as stochastic processes in \eleca. 
 
Each particle of the cascade is followed during the propagation until its energy reaches a threshold which can be set by the user. In the present work, we consider an energy threshold of $10^{16}$~eV, if not differently specified. The cosmology is given by the $\Lambda$-cold dark matter model with $H_0= 70.4$~km/s/Mpc, $\Omega_{M}=0.272$ and  $\Omega_{\Lambda}=0.728$. Moreover, in order to reduce the computational time, an option can be activated to propagate only particles with a chance larger than 1\% to have at least one electromagnetic particle reaching the Earth (see section~\ref{sect:maxdist}).

\begin{figure}[!t]
\includegraphics[width=0.5\textwidth]{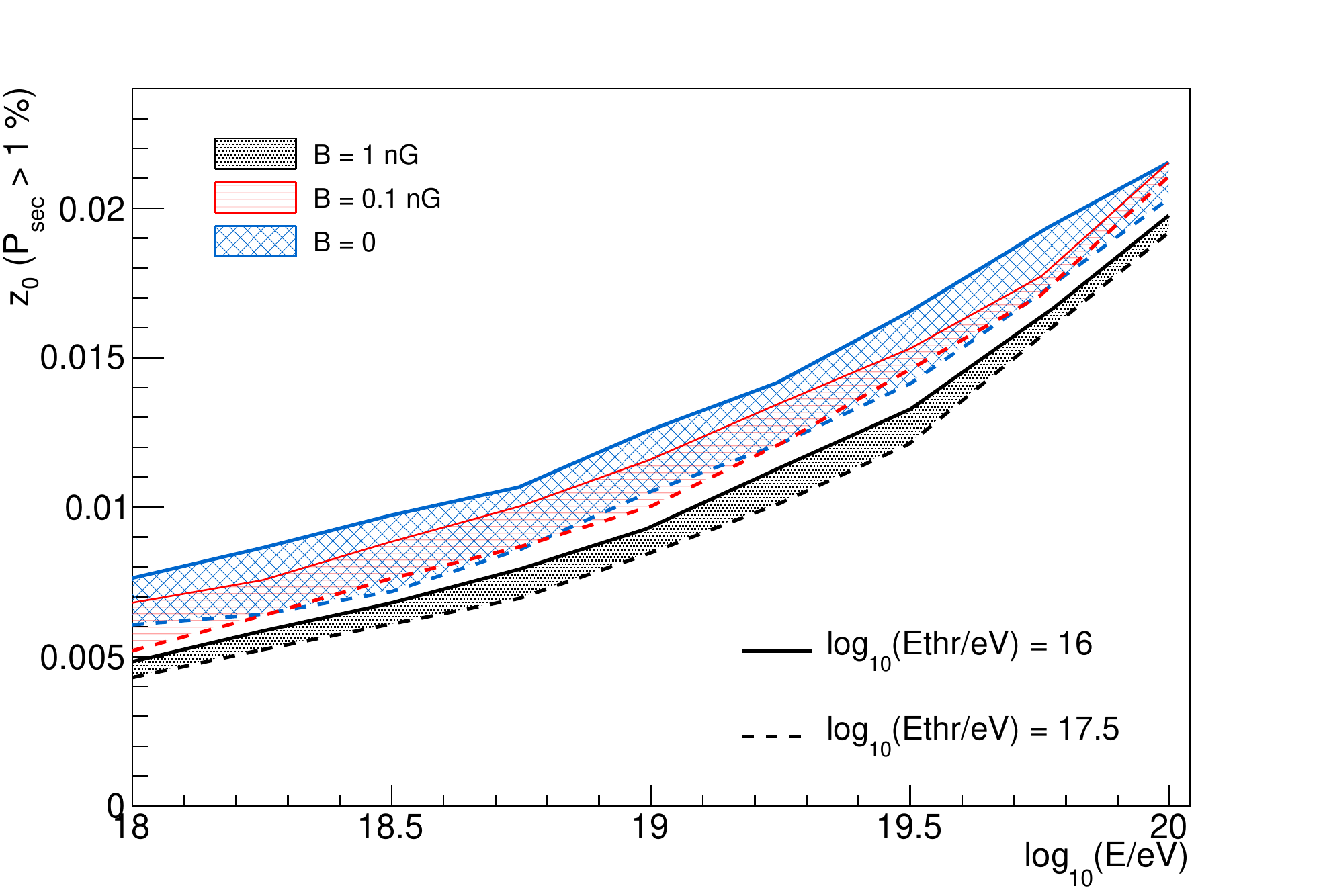}
\caption{Maximum distance $z_{\rm{max}}$ corresponding to a probability of observing at least one secondary particle above a given energy threshol $E_{\rm{thr}}$ and for a given initial configuration (E$_{0}$,z$_{0}$, B).}
\label{fig:maxdist}
\end{figure}

\section{Observation probability for secondary photons}\label{sect:maxdist}

In order to gain further insights into electromagnetic cascade propagation, we estimate the probability of observing a secondary photon at Earth versus the energy (E$_{0}$) and the redshift (z$_{0}$) of the photon at the production site. 
More specifically we define a probability $P_{\rm{sec}}$ of observing at least one $\gamma$ above the energy threshold E$_{\rm{thr}}$ for a given initial configuration (E$_{0}$, z$_0$). 
In Fig.~\ref{fig:maxdist} we show the distance $z_{\rm{max}}$ corresponding to P$_{\rm{sec}} > $ 1\% versus the energy E$_{0}$. Different magnetic fields configurations are shown as bands, where each band is delimited by the two energy thresholds (i.e., $10^{17.5}$ and $10^{16}$~eV). As shown in figure, the secondaries of cascades induced by photons with energy up to $10^{20}$~eV are unlikely to reach the Earth from distances larger than about 80~Mpc, assuming an energy threshold of $10^{16}$~eV. In the next section such an effect is convolved with the propagation of UHE nuclei and their production of UHE photons because of the interactions with the EBR.
From a practical point of view, we use the results in Fig.~\ref{fig:maxdist} to define a maximum distance $z_{max}$ beyond which photon propagation does not significantly affect the photon flux at the Earth. A parameterization of $z_{\rm{max}}$ for the most conservative case (B=0) can be used to speed up the simulations by neglecting the cascade development induced by photons beyond the maximum distance z$_{\rm{max}}$(E). We checked that the effect of applying such a maximum distance cut is negligible on the flux of secondary particles at Earth. As expected, we found that the observed spectrum is affected by less than 1\%, independently of the energy of particles at the observer.

\section{GZK photons}
In this section we focus on one relevant mechanism responsible for the production of UHE photons, i.e., the GZK effect. In fact, protons with energy above $\approx50$~EeV and heavier nuclei with Lorentz factor larger than $10^{11}$ \cite{allard2005uhe,Hooper:2006tn,globus2008propagation,hooper2011cosmogenic,DeDomenico:2013psa}, while interacting with relic photons mainly experience baryonic resonances which produce neutral mesons, mainly pions, that in turn quickly decay to two $\gamma$s. It is worth remarking that the number of UHE photons produced during nuclei propagation, as well as their production sites and their expected energy spectrum at Earth, depends on the assumptions about the sources. More specifically, the injection mechanism of nuclei is not known: the energy spectrum at source is assumed to follow a power law $E^{-\gamma}$ with a cut-off at E$_{\rm{max}}$ that can be chosen as a step function: 
\begin{equation}
E^{-\gamma}\Theta(E - E_{\rm{max}})
\label{eq:1}
\end{equation}
or an exponential decrease:
\begin{equation}
E^{-\gamma}e^{-E/E_{\rm{max}}}. 
\label{eq:2}
\end{equation}

Moreover, the chemical composition at the source, as well as the relative flux normalization for each primary type,  the true distribution of sources and their redshift evolution are still unknown. The existing models of EBR and extragalactic magnetic fields (r.m.s. strength and coherence length of the turbulent component) provide additional astrophysical and cosmological observables to vary in simulations. 

The experimental observation (or non observation) of a photon flux compatible with the fluxes predicted by means of detailed simulations can be used to get insights into the sources and the extragalactic magnetic fields. 

Here, we consider several possible scenarios, varying the source injection, evolution and maximum acceleration, as well as nuclei abundances and intervening EGMF. The nuclei propagation is performed with \hermes, a simulator based on a Monte Carlo approach~\cite{DeDomenico:2013psa,DeDomenico:2013doa}. The produced GZK photons are successively propagated with \eleca\footnote{It is worth remarking that \eleca\, is a stand-alone code that can be used to propagate photons regardless of the specific nuclei propagator.}.

In particular, we consider sources homogeneously distributed up to $z_{max}=2$ and the two injection mechanisms in Eq.~(\ref{eq:1}) and Eq.~(\ref{eq:2}). Different nuclear species at the source, namely proton, helium, oxygen and iron, are also considered. The impact of the source evolution with redshift is also investigated.

\subsection{Horizons for GZK photons}\label{sect:horizons}

The GZK horizon \cite{Greisen:1966jv,Zatsepin:1966jv} is generally introduced to provide an estimation of the maximum distance within which we expect that 90\% of UHECRs -- specifically, light or heavy nuclei -- will reach the Earth above a given energy threshold~\cite{deligny2004magnetic,parizot2004gzk,harari2006ultrahigh,kachelriess2008gzk,de2013influence}. 

Here, we estimate the probability of observing GZK photons under different astrophysical hypothesis. Even if GZK is typically referred to proton interaction models, it is commonly used also for nuclei interacting on CMB. Here, we define the ``$\gamma-$GZK horizon'' in a similar spirit, provided that it refers to cosmogenic gamma horizon. Following~\cite{de2013influence} for the case of nuclei, we define $\omega_{GZK}(z;E_{\rm{thr}},A)$ as the probability that a secondary photon reaches the distance of the Earth with an energy equal to or larger than $E_{\rm{thr}}$ if a \emph{single source} at distance $z$ injects nuclear primaries with mass $A$. Therefore, the $\gamma-$GZK horizon is derived from $\omega_{GZK}(z;E_{\rm{thr}},A)$ by considering the contribution of \emph{all sources} between $z = 0$ and $z = 2$ to the flux. We define the $\gamma-$GZK horizon -- for photons above energy threshold $E_{\rm{thr}}$ produced by nuclear species $A$ -- as the maximum distance $D_{f}(A)$ within which a certain fraction $f$ of the total photon flux reaches the Earth. 

\begin{figure}[!t]
\hspace{-0.4cm}
\includegraphics[width=0.50\textwidth]{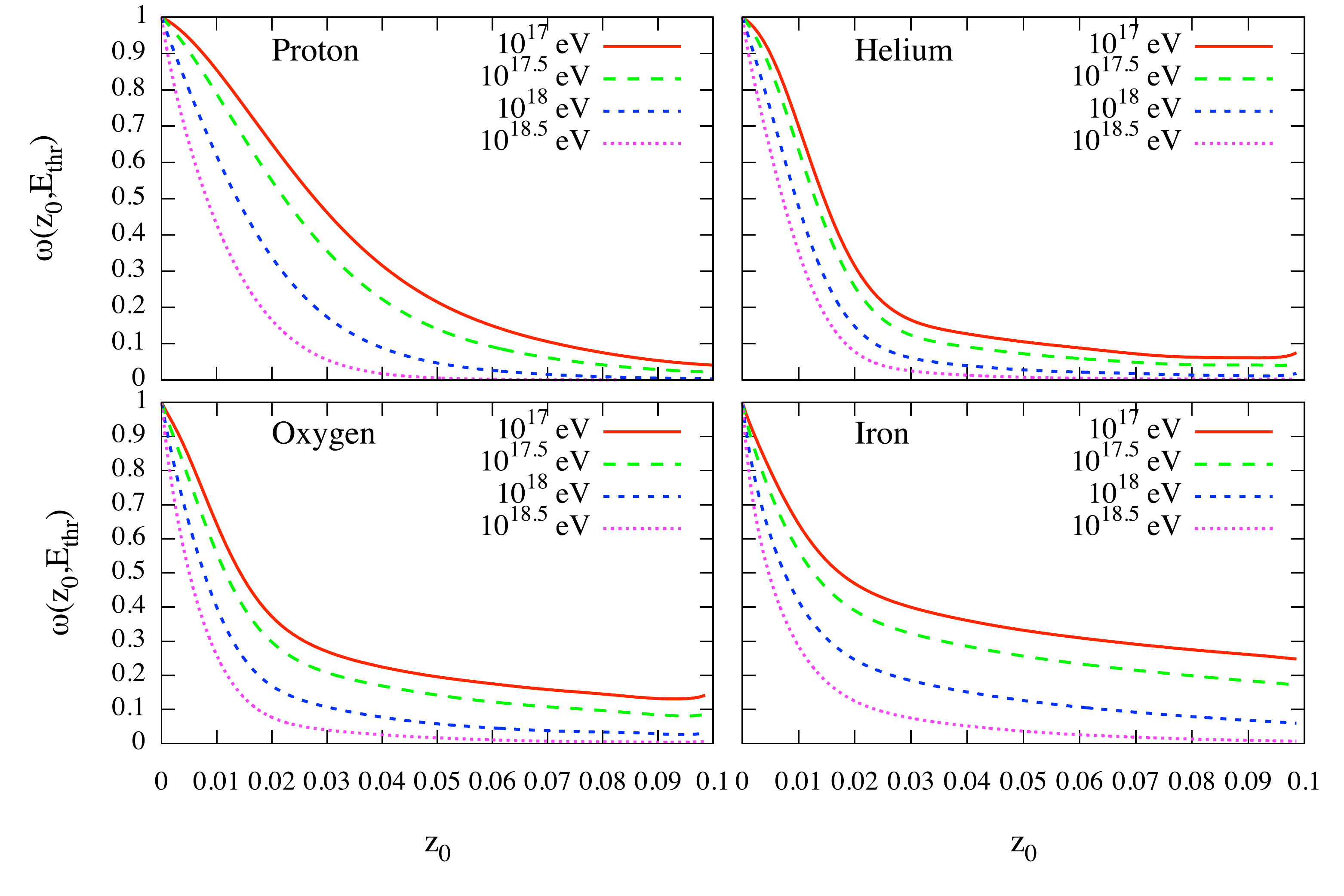}
\caption{Probability of observing at the distance of the Earth at least one photon above the energy threshold indicated in the legend. The curves correspond to scenarios with different nuclei injected at the source with redshift $z_{0}$.}
\label{fig:gzk_omega}
\end{figure}

In Fig.~\ref{fig:gzk_omega} we plot the probability $\omega_{GZK}(z;E_{\rm{thr}})$ for different energy thresholds, considering an homogeneous distribution of equal-intrinsic-luminosity sources uniformly distributed up to $z = 2$. We consider different scenarios with different mass composition, with source injecting nuclei following a power-law spectrum with spectral index 2.3 and energy cutoff $Z\times 10^{21}$ eV. The nuclei may produce photons during their propagation (namely ``primary photons''), as discussed in the previous section, which propagate through the Universe, eventually inducing an electromagnetic cascade. Possible point sources directly producing photons (e.g.~ GRB~\cite{PhysRevLett.103.081102,2012ApJ...745L..16M}) are not considered here and will be part of a dedicated study. 
Hence, we calculate $\omega_{GZK}(z;E_{\rm{thr}})$ as the probability that at least \emph{one} photon (being the initial GZK photon or a photon produced in the electromagnetic cascading) reaches the distance of the Earth above a given energy, as a function of the distance $z$. 

\begin{figure}[!t]
\hspace{-0.4cm}
\includegraphics[width=0.50\textwidth]{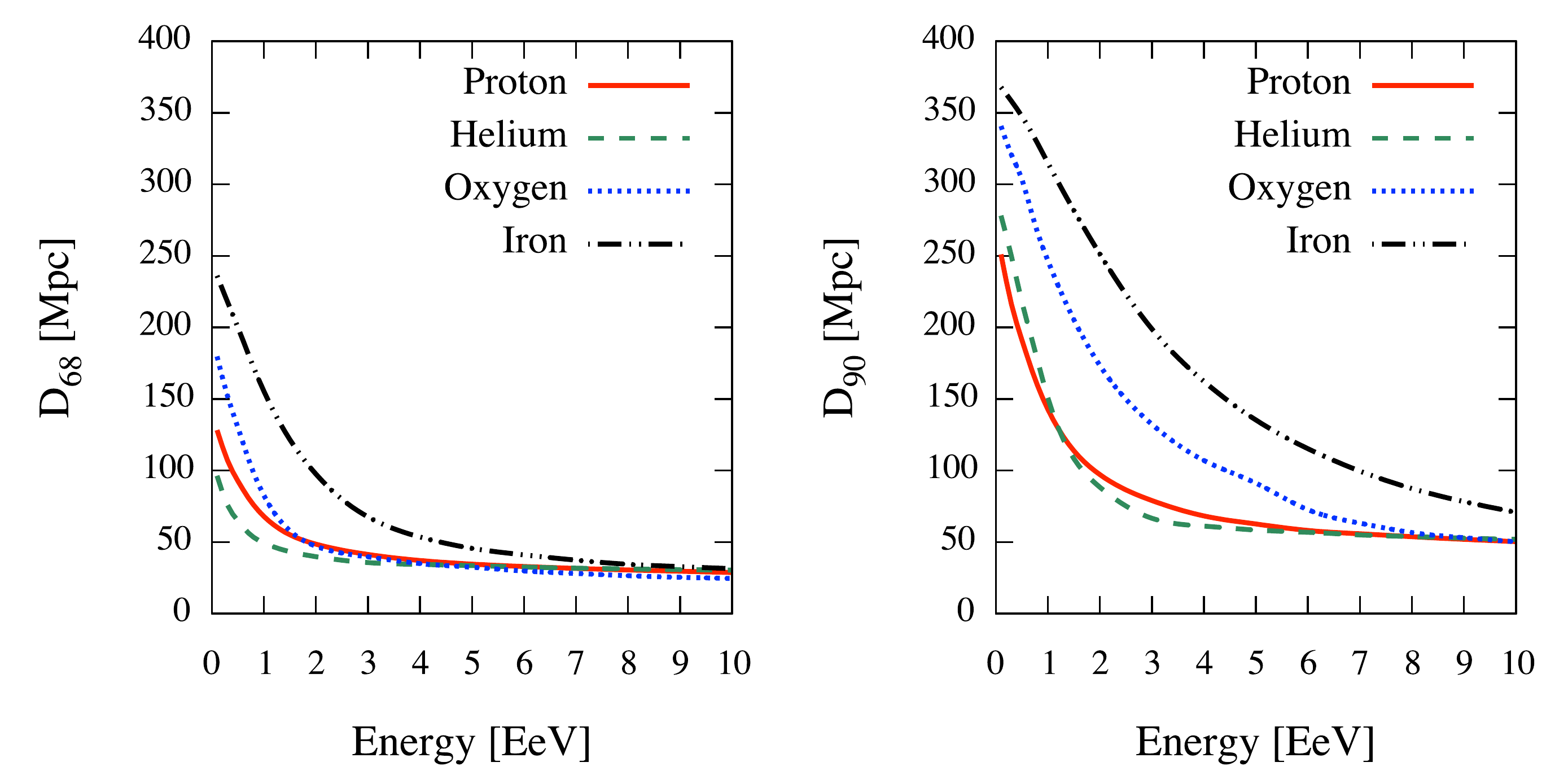}
\caption{The $\gamma-$GZK horizon $D_{68}$ (left) and $D_{90}$ (right) is shown for UHE photons observed at Earth with energy above the one reported in the $x$-axis (see the text for further detail). Photons produced during the propagation of nuclei injected at source with mass $A$ and power-law spectrum $E^{-2.3}$ are considered.}
\label{fig:gzk_horizon}
\end{figure}

We show in Fig.~\ref{fig:gzk_horizon} the $\gamma-$GZK horizon in the case of four different injected nuclear species (i.e., H, He, O and Fe) at the source, as a function of the energy threshold. Two representative values of the fraction $f$ are shown: D$_{68}$ (left), where $f=$68\% and D$_{90}$ (right) where $f=$90\%.

\begin{figure}[!t]
\hspace{-0.4cm}
\includegraphics[width=0.50\textwidth]{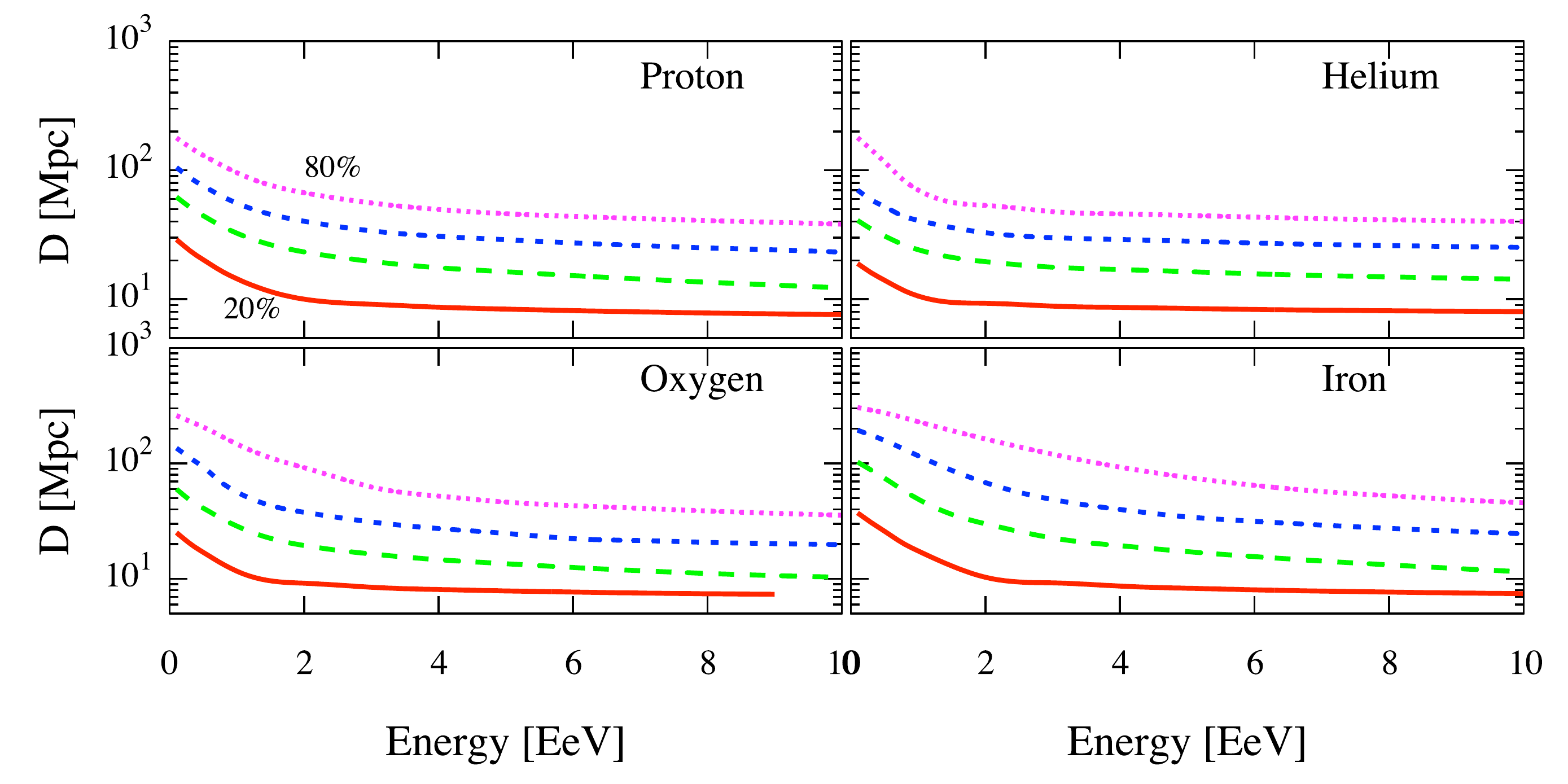}
\caption{Dependence of the $\gamma-$GZK horizon on the energy threshold at Earth, the injected nuclear composition and the fraction $f$ of the integral photon flux ($f = 0.2, 0.4, 0.6$ and $0.8$).}
\label{fig:gzk_horizon_evolution}
\end{figure}

Fig.~\ref{fig:gzk_horizon_evolution} shows the dependence of the $\gamma-$GZK horizon for photons produced by  different nuclei primaries and for different choices of the fraction $f$. 

According to these plots, photons with energies above a given threshold are expected to be produced by sources located in the nearby universe, up to 250~Mpc in the case of injected protons or up to $\sim$350~Mpc in the case of injected iron. This result shows that the observation of sources of UHE nuclei, by means of UHE photons, is limited to the nearby Universe. 
It is worth mentioning that, the possibility of doing astronomy with UHE photons, in the case of future observations, is limited by factors altering the arrival directions of such photons, in particular by the intervening magnetic fields on the charged component of the electromagnetic cascade.
These deflection depend on the EGMF strength and on the source distance and on the development of the cascade. For GZK photons, the total deflection of the observed photons would add to the deflection experienced by the primary nuclei. The combination of the angular separation between a detected photon and possible sources, combined with the probability of detecting photons from a nuclear source at distance $z$, can be hypothetically used to constrain the nature of the primary CRs and the intensity of the magnetic fields they traverse. 
The deflection of the electromagnetic cascade in the EGMF is a well known interesting topic especially in the gamma-rays energy range (from 100 MeV to TeV)~\cite{Ahlers:2011sd,Aharonian:1994,Elyiv:2009bx,d'Avezac:2007sg} where the temporal and the angular characteristics of the electromagnetic cascades from known sources are used to set limits on the strength of the EGMF (see for example~\cite{1995Natur.374.430P,Ando:2010rb,2041-8205-771-2-L42,Tavecchio:2010mk,Neronov:1900zz,Dolag:2010ni}). 

\subsection{Photon fluxes}\label{sect:fluxes}

In this section we briefly discuss how different astrophysical scenarios influence the flux at Earth. 
The normalization of the fluxes is obtained by scaling the corresponding all-nuclei flux at $10^{18.85}$~eV to the most recent value measured by the Pierre Auger Observatory~\cite{Auger:icrc2013}.

\begin{figure}[!t]
\hspace{-0.4cm}
\includegraphics[width=0.50\textwidth]{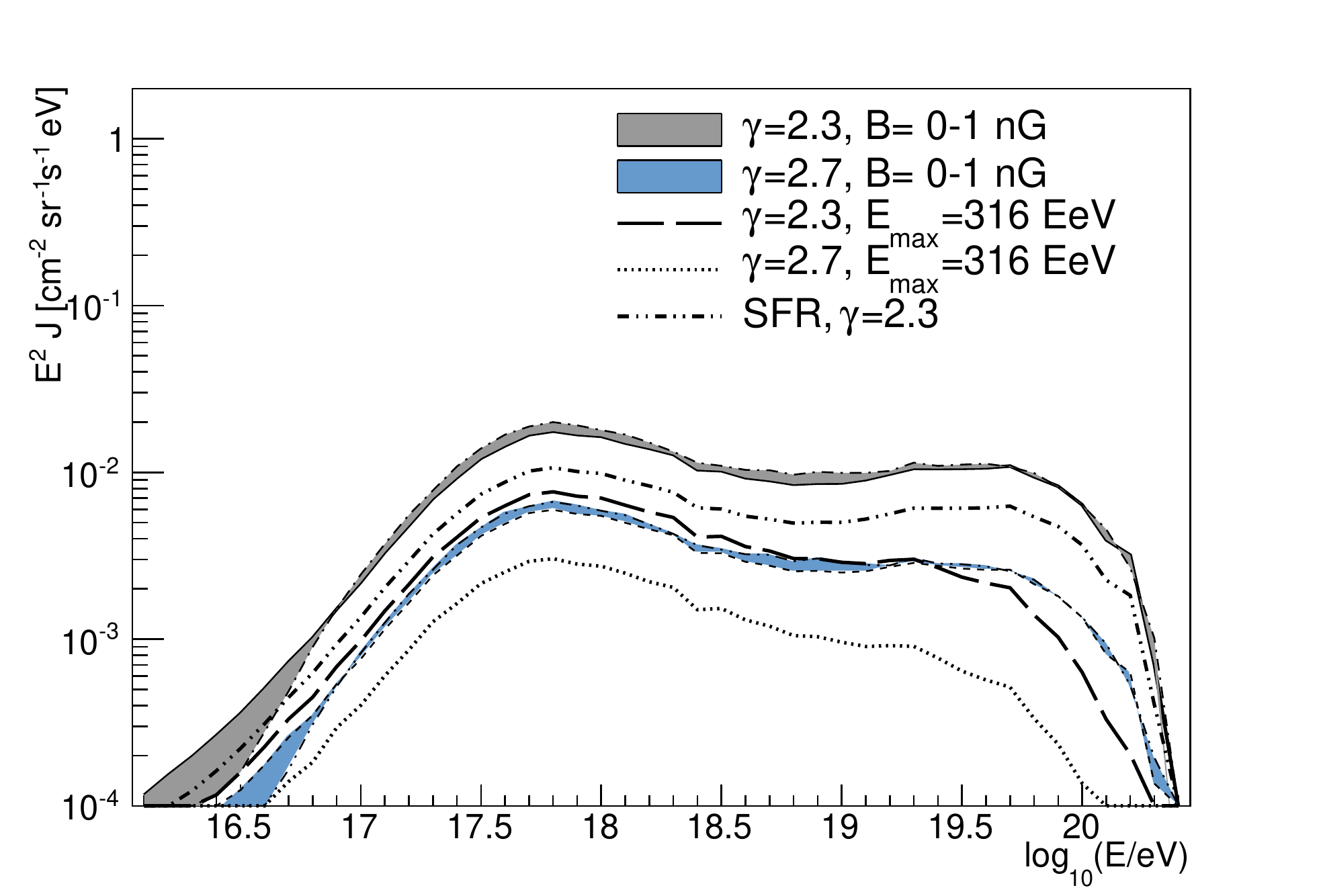}
\caption{Differential photon flux predicted at Earth assuming different scenarios at the source (i.e., varying spectral index, energy cut-off and redshift evolution) and intervening extragalactic magnetic fields. }
\label{fig:gzk_mixed}
\end{figure}

The influence of the source evolution with redshift and of the intergalactic magnetic fields is shown in Fig.~\ref{fig:gzk_mixed}. In particular, we compared the scenario not accounting for the evolution -- as assumed so far -- against a scenario where evolution follows star formation rate (SFR). Other redshift evolutions can be considered as well with our simulator, and their effect is to alter the intensity of the flux without substantial changes to the shape of the observed spectrum. We also show the flux modification due to the absence ($B=0$) or presence ($B=1$~nG) of an EGMF acting on the photon cascade. The flux is mainly influenced at lower energy, where the deflection of the electromagnetic component is more sensitive to the magnetic field. It is worth mentioning here that the structured magnetic fields in the nearby Universe (within $\approx100$~Mpc) are not included in our simulations.  

\begin{figure}[!t]
\hspace{-0.4cm}
\includegraphics[width=0.50\textwidth]{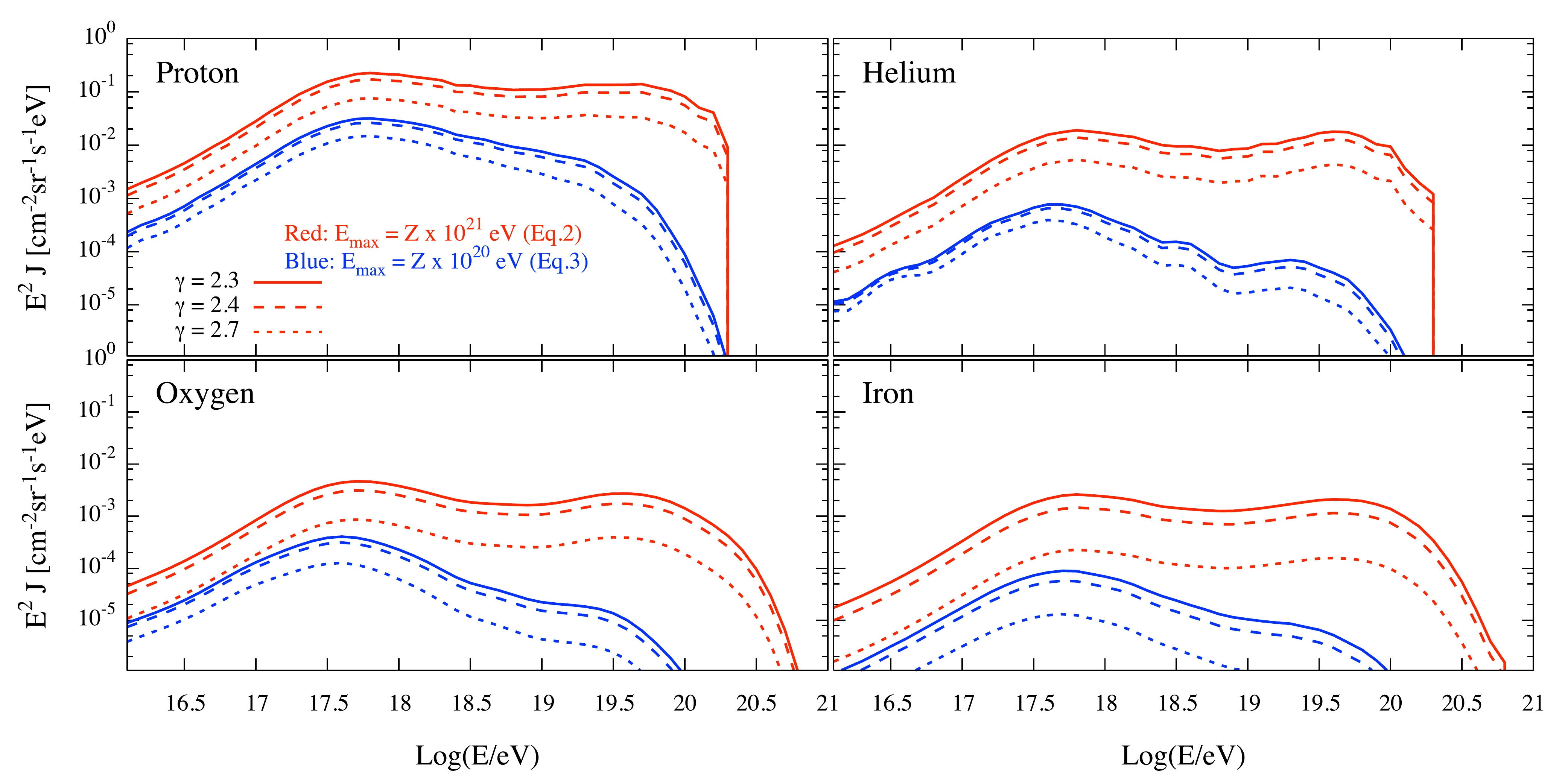}
\caption{Differential photon flux for different spectral indices (see legend) and for the four nuclei considered in this work (proton to iron from top-left to bottom-right).}
\label{fig:gzk_Gdep}
\end{figure}

Fig.~\ref{fig:gzk_Gdep} shows the impact of the injection mechanism of UHECRs, varying both the spectral index and the maximum energy at the source. 
Since this study is devoted to photon propagation, we have not performed an exhaustive search for the value of $\gamma$ which optimizes the fit of the simulated all-particle spectrum to the experimental measurements. The maximum injection energy was varied and the result is also shown in the same figure with different colors. In particular, we considered the case of an energy spectrum with a sharp cut-off at $E_{\rm{max}}=Z\times10^{21}$~eV (see Eq.~(~\ref{eq:1})) (black), and an exponential suppression (see Eq.~(\ref{eq:2})) with $E_{\rm{max}}=Z\times10^{20.5}$~eV. A steeper spectrum (or a limited power at the source) significantly reduces the photon flux at Earth because of the suppression of the highest energy primaries, which mostly contributes to the spectrum especially above 10$^{19}$~eV. 
Therefore, the energy range above this threshold is of particular interest to exploit - and eventually exclude - the scenarios with a limited maximum energy at the source.

\begin{figure}[!t]
\hspace{-0.4cm}
\includegraphics[width=0.50\textwidth]{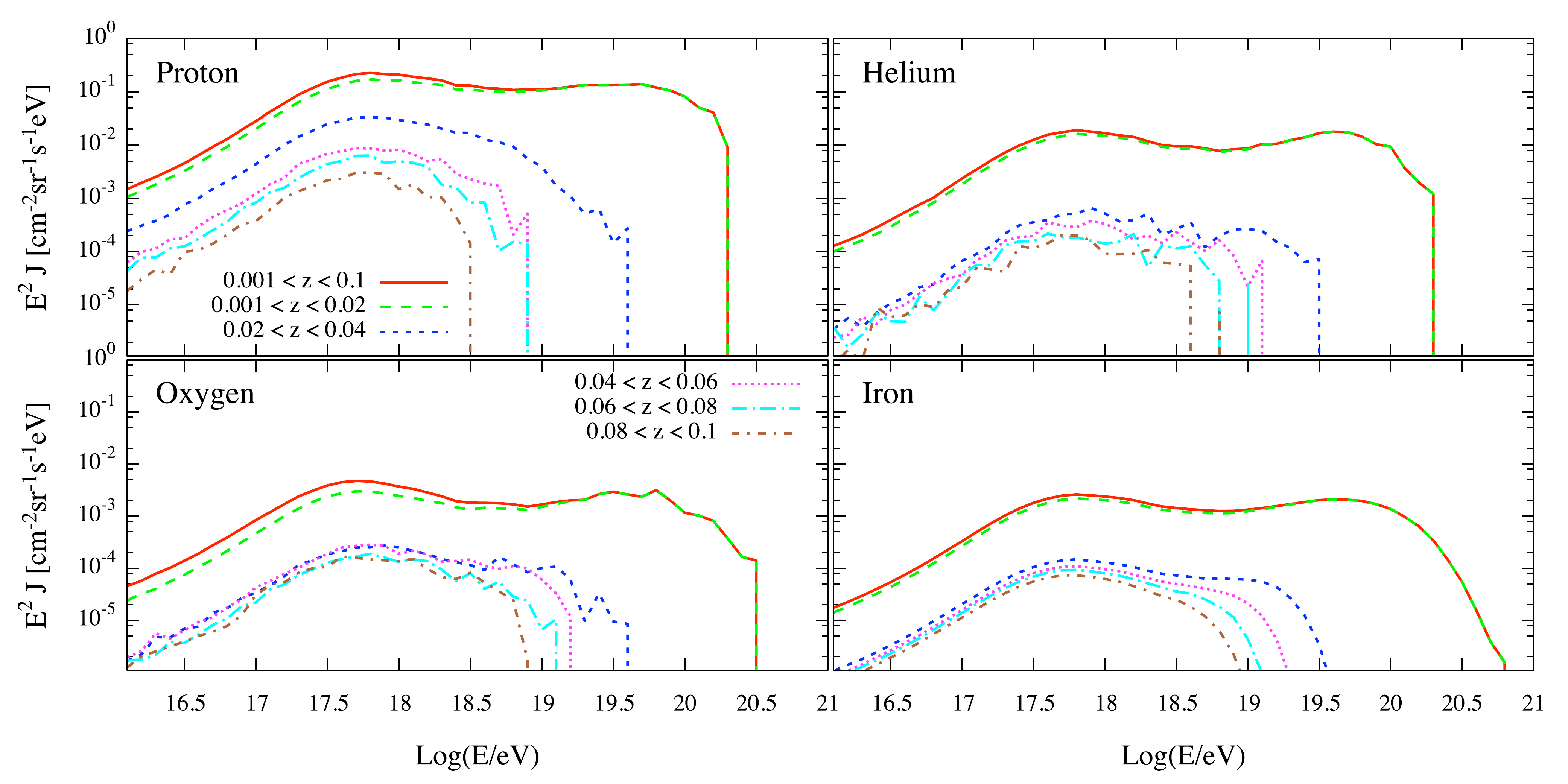}
\caption{Differential photon flux for distances of the sources of UHECRs. Panels as in the previous figure.}
\label{fig:gzk_Rdep}
\end{figure}

In Fig.~\ref{fig:gzk_Rdep}, we investigate the impact of the source distance on the flux. In this study, all the sources are assumed to be uniformly distributed and with equal luminosity. We are only interested in sources within $z = 0.1$, since the flux contribution above this distance is expected to be sub-dominant. Each panel of the figure corresponds to a primary mass (lighter to heavier reading from top-left to bottom-right). The total flux (up to $z = 0.1$) is shown as a solid line whereas the contributions from various distance shells are shown separately as dashed lines. The total flux is normalized, as discussed before, to the Auger measured flux. 
The closest sources (within $z<0.02$) are responsible for the highest energy photon flux while, at energies below $10^{19}$~eV, a contribution of about 10\% is due to farther sources as a result of the cascade development. This contribution is even smaller for heavier nuclei as a reflection of the different mean interaction length for photon production and of the energy spectrum of the produced $\gamma-GZK$.
From the results shown before, the observation of a ``significant'' photon flux at energies above $10^{19}$~eV is thus a strong signature of a UHECR source in the nearby universe and with high E$_{\rm{max}}$. In the case of nearby sources, the combined information with anisotropy in the arrival direction of charged cosmic rays would help to distinguish between heavy and light composition scenarios. 

\begin{figure}[!t]
\hspace{-0.4cm}
\centering
\includegraphics[width=0.40\textwidth]{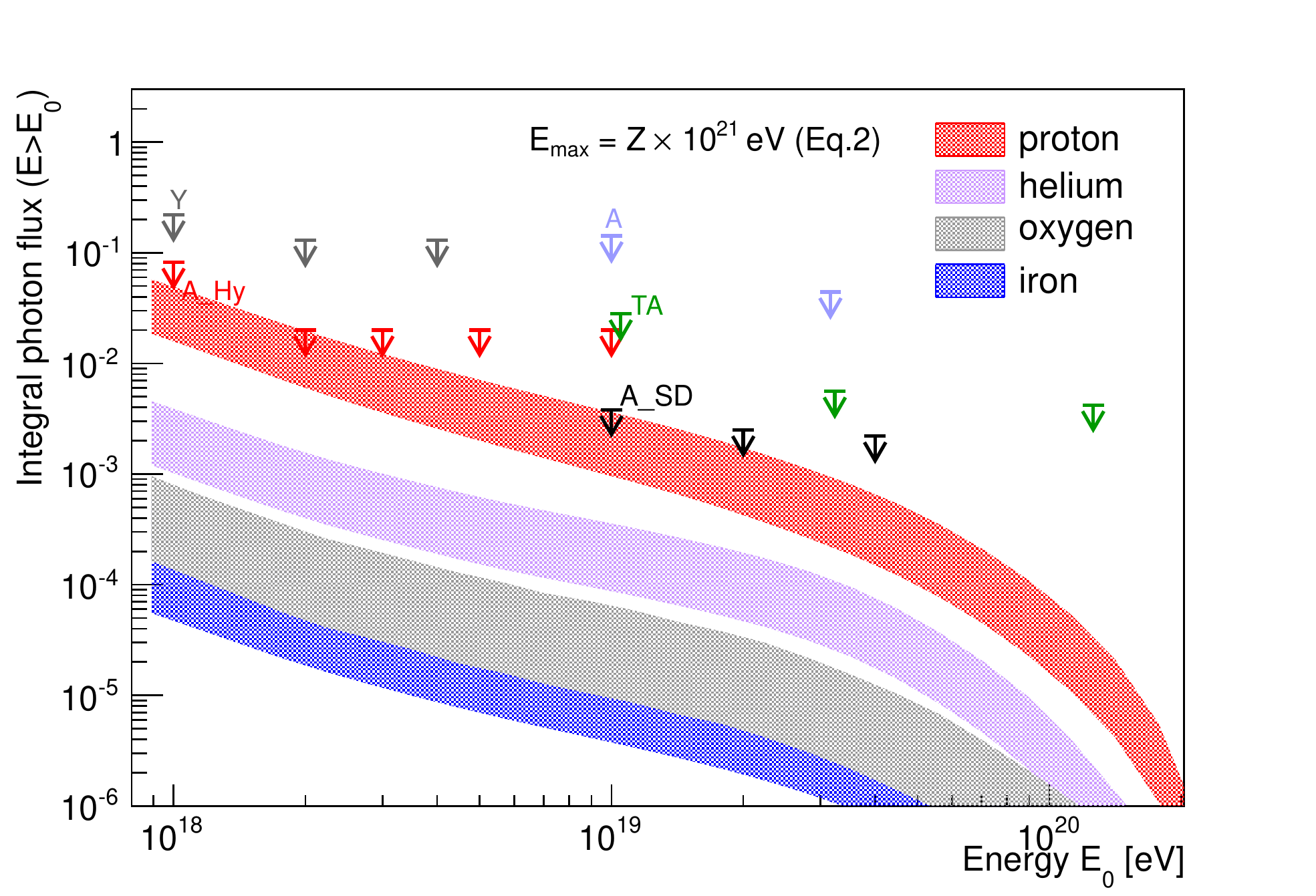}
\caption{Integral photon flux above the energy threshold E$_{0}$ as a function of energy, in the scenarios with sources accelerating proton (red), helium (violet), oxygen (gray) and iron (blue) up to a maximum energy 10$^{21}$~eV assuming a power law spectrum with a cut-off at E$_{\rm{max}} = Z\times10^{21}$~eV. The shaded areas are due to the different spectral indices assumptions as in Fig.~\ref{fig:gzk_Gdep}. The experimental results are also shown for the Pierre Auger Observatory (A$_{Hyb}$~\cite{Settimo:2011zz}, A$_{SD}$~\cite{Abraham:2009qb}), AGASA (A)~\cite{AgasaLimits}, Yakutsk~\cite{YakutskLimits} and Telescope Array~\cite{TelescopeArray}.}
\label{fig:integralfluxes}
\end{figure}

Upper limits on the photon flux (or on the fraction of photons in the all-particle flux) have been placed with ground-based experiments. A plot of the integral photon fluxes as a function of the minimum integration energy is shown in Fig.~\ref{fig:integralfluxes} for one of the scenarios discussed before, namely the one where the spectral flux has a sharp cut-off at E$_{max} = Z\times 10^{21}$~eV. We consider the case of different primary nuclei and we emphasize the effect of the different spectral indices (delimited by the shaded bands). For comparison, the current experimental limits derived by the Pierre Auger Observatory~\cite{Settimo:2011zz,Abraham:2009qb}, AGASA~\cite{AgasaLimits}, Yakutsk~\cite{YakutskLimits} and Telescope Array~\cite{TelescopeArray} are shown. This plot is given for illustrative purpose since the identification of optimal astrophysical scenarios to describe the current observations (energy spectrum, mass composition and photon flux limits) is out of the scope of the present work and will be the subject of successive studies. 

\section{Conclusions and outlook}

\eleca\,, simulating the development of electromagnetic cascades initiated by UHE photons and e$^+$/e$^-$, is presented in this paper. It is a C++ Monte Carlo code with high modular structure that can be easily combined with external simulators developed, for instance, to propagate UHE nuclei (here \hermes). The flexibility of our simulator allows us to perform several types of studies. 

In this work, we focused on the prediction of GZK photon fluxes expected at Earth in different astrophysical scenarios. More specifically, we show our results in the energy range above $10^{16.5}$~eV. Similarly to the case of nuclei, the fluxes at Earth might differ by several orders of magnitude depending on the chosen environmental configurations (e.g., extragalactic background radiation, magnetic fields, \emph{etc}), source distribution and injection mechanisms, or chemical composition. 

To identify the possible origin of an hypothetical observed high energy photon, we have introduced the concept of ``$\gamma-$GZK horizon'' to derive an analog of the GZK horizon used for nuclei. We have defined the $\gamma-$GZK horizon -- for photons above energy threshold $E_{\rm{thr}}$ produced by nuclear species $A$ -- as the maximum distance $D_{f}(A)$ within which a certain fraction $f$ of the total photon flux reaches the Earth. First, we have estimated the probability that a secondary photon reaches the Earth with an energy equal or larger than a given threshold, under the assumption that a \emph{single source} at distance $z$ injected nuclear primaries with mass $A$. The $\gamma-$GZK horizon was successively derived by considering the contribution of \emph{all sources} between $z = 0$ and $z = 2$ to the flux. Our findings show, for instance, that if a photon of 1~EeV is observed, it is very likely (90\% probability) that it has been generated by interactions between relic photons and a proton (or an Helium nucleus) injected within 150~Mpc, or an iron nucleus injected within 300~Mpc. For an energy larger than 10~EeV this distance reduces to 50~Mpc, regardless for the injected composition.

On the other hand, we have shown that intervening magnetic field have an impact on the observed photon flux only at energies lower than $10^{17.5}$~eV and that the value of the maximum injection energy might further reduce the flux above $10^{20}$~eV. Therefore, the range between these values of energy is the most suitable for the detection of high energy photons. In particular, the flux observed at different energies may provide information on the energy spectrum at the sources and their distances. In particular, we have shown that the energy range above $10^{19}$~eV, is the most sensitive to the maximum energy of the injection spectra and is the one interested exclusively by sources within few tens of Mpc. In a scenario without close and highly energetic sources, the lower energy range is however the most relevant for the detection. 
From the experimental point of view, the largest ground-array detectors, in particular the Pierre Auger Observatory~\cite{Settimo:2011zz}, are reaching the sensitivity to explore the region where a flux of $\gamma-GZK$ can be experimentally observed. 
Although the photon flux depends on several astrophysical factors, our numerical predictions combined with future experimental observations (or non-observations) of UHE photons can help to constrain some of these factors, thus reducing the size of the parameter space. 

Finally, further improvements, a 3D version with a more realistic treatment of extragalactic magnetic fields, new environment models (e.g. radio background, extra-galactic medium) and optimization for the propagation of photons at the TeV scale are in progress and will be the subject of later studies.

\section*{Acknowledgments}
The authors are grateful to Markus Risse and Haris Lyberis for useful discussions.\\
The work of M.S., made in the ILP LABEX (under reference ANR-10-LABX-63), is supported by French state funds managed by the ANR within the Investissements d’Avenir programme under reference ANR-11-IDEX-0004-02.

\bibliographystyle{mprsty}

\end{document}